\documentclass[prl,twocolumn,showpacs,floatfix]{revtex4}
\usepackage{graphicx}

\usepackage{amssymb}
\usepackage{amsbsy}
\usepackage{psfrag}

\begin{document}

\title{Density-PDFs and Lagrangian Statistics of highly compressible
  Turbulence}

\author{Christoph \surname{Beetz}}
\author{Christian \surname{Schwarz}}
\author{J\"urgen \surname{Dreher}}
\author{Rainer \surname{Grauer}}
\affiliation{Theoretische Physik I, Ruhr-Universit\"at Bochum, Germany}

\date{\today}

\begin{abstract}
  \noindent
  We report on probability-density-functions (PDF) of the mass density
  in numerical simulations of highly compressible hydrodynamic flows
  and the corresponding structure formation of Lagrangian particles
  advected by the flows. Numerical simulations were performed with
  $512^3$ collocation points and 2 million tracer particles integrated
  over several dynamical times. We propose a connection between the
  PDF of the Lagrangian tracer particles and the predicted log-normal
  distribution of the density fluctuations in isothermal systems.
\end{abstract}


\maketitle

\noindent
\textit{Introduction:}
Compressible fluid and plasma turbulence is believed to be of
fundamental importance for understanding relevant processes in
astrophysics. A typical example is the effect of turbulence on the
star formation rate in dense molecular clouds in the interstellar
medium \cite{padoan-nordlund:2002}. The effect of turbulence is
manifested in the highly intermittent nature of density fluctuations
in supersonic flows which naturally has an enormous impact on the
local conditions triggering star formation.

Traditionally, turbulence is studied in the Eulerian description,
where the temporal evolution of the fluid quantities is followed at
fixed spatial locations. However, in the last 10 years the Lagrangian
description, which describes the motion of fluid particles, has
undergone a rapid development.  This is mainly due to advanced
experimental techniques introduced in Ris\o \ \cite{ott-mann:2000},
Cornell \cite{porta-voth-etal:2000, voth-porta-etal:2001,
  porta-voth-etal:2001} and Lyon \cite{mordant-metz-etal:2001} for
studying the statistics of passive tracer particles in incompressible
turbulent fluids. Both, experimental and numerical
\cite{biferale-bofetta-etal:2004, homann-grauer-etal:2007} studies
highlight the importance of singular structures like vortex tubes and
sheets for the intermittent statistics of incompressible flows.
Lagrangian statistics is not only interesting for obtaining a deeper
understanding of the influence of typical coherent or nearly-singular
structures in the flow but also of fundamental importance for
understanding mixing, clustering and diffusion properties of turbulent
astrophysical fluid and plasma flows.

In this Letter, we will use the Lagrangian description of passive
tracer particles in highly compressible flows. Due to the
compressibility, the situation is very different from the
incompressible case. Clustering of passive tracer particles will
reflect the strong density fluctuations. This effect does not appear
in the incompressible case. However, the incompressible case will
serve as a guideline to obtain a mapping between the PDF of density
fluctuations and the PDF of the particle distribution.

The goal of the paper is threefold: The main result is an analytical
expression for the spatial distribution of the Lagrangian tracers.
This result will be used to confirm and quantify the relation between
the standard deviation of the Eulerian density fluctuations and the
mean Mach number. And last, the analytical description helps to define
a precise structuring time for the clustering of the Lagrangian
particles which turns out to be significantly shorter than the
dynamical time.

\noindent
\textit{Numerical Methods:}
To follow this approach in the examination of compressible fluid
turbulence we solved the isothermal Euler equations for density $\rho$
and momentum density $\mathbf{u}=\rho\mathbf{v}$
\begin{eqnarray}
  \partial_t \rho + \nabla\cdot \mathbf{u} &=& 0 \nonumber \\
  \partial_t \mathbf{u} + \nabla\cdot(\frac{\mathbf{uu}}{\rho})
  &=&-\nabla P + \rho \mathbf{k}
  \label{equ:euler}
\end{eqnarray}
with $P \sim \rho$ using direct numerical simulations in real space,
and realized Lagrangian measurements with tracer particles, which are
advected by the simulated flow.  We performed three runs of stationary
flows (a, b and c) with mean r.m.s.\ Mach numbers $\tilde M$ of $0.4$,
$1.4$ and $4.6$, respectively.

Equations (\ref{equ:euler}) were solved in the
\mbox{\em{\bfseries{racoon}}} environment for hyperbolic differential
equations \cite{dreher-grauer:2005}. It utilizes a third order
semi-discrete central scheme \cite{kurganov-levy:2000}. Integration is
done by a third order strongly stable Runge-Kutta scheme
\cite{shu-osher:1988}. A 3-dimensional cube of length $L=2\pi$ was
discretized in physical space with $\mathcal{N}^3= 512^3$ collocation
points and periodic boundary conditions. The simulation was started
with an Orszag-Tang like initial condition and a homogeneous density.
The simulations were run for several dynamical times $t_{dyn}=L/2M$ to
let the system evolve into a turbulent state. The mean energy input
was realized by a temporal delta correlated stirring force with zero
mean momentum. Forcing was applied in a shell $k\in[1,2]$ in Fourier
space, such that it acts only on scales comparable to the entire box.
The amplitudes of the Fourier modes were adjusted in order to achieve
stationarity at various mean Mach numbers.  Then the mean energy input
was kept constant for each run.

In this evolved turbulent system we inserted \mbox{$N=2\,000\,000$}
particles. They were randomly distributed over the computational
domain and then advected by the flow field $\mathbf{v}(\mathbf{x}) =
\frac{\mathbf{u}(\mathbf{x})}{\rho(\mathbf{x})}$.  Particles were
integrated using the same third-order Runga-Kutta scheme as for the
fields.  The field values at the particle positions $\mathbf{x}$ were
evaluated by a linear interpolation, since the density has to be
positive even in regions where strong shocks cause steep gradients
close to flat areas. Higher order interpolators here tend to
undershoot which leads to artificial particle reflection.  After
another dynamical time $t_{dyn}$ to assure that the particle
statistics has reached a stationary state, data were analyzed both for
Eulerian and the Lagrangian statistics.

\begin{figure}[t]
  \centering
  \includegraphics[width=.95\linewidth]{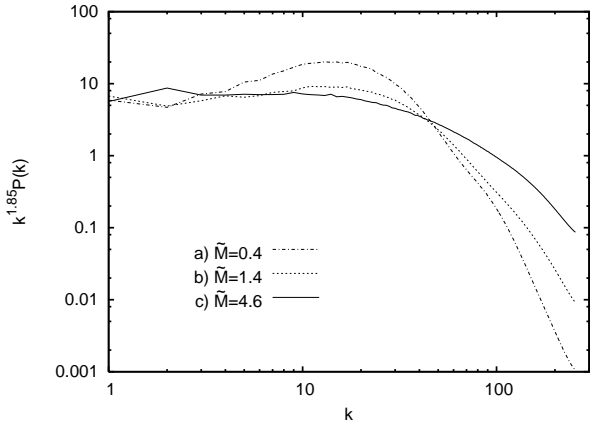}
  \caption{Compensated velocity spectra}
  \label{fig:spectra}
\end{figure}

\begin{figure}[t]
  \centering
  \includegraphics[width=.95\linewidth]{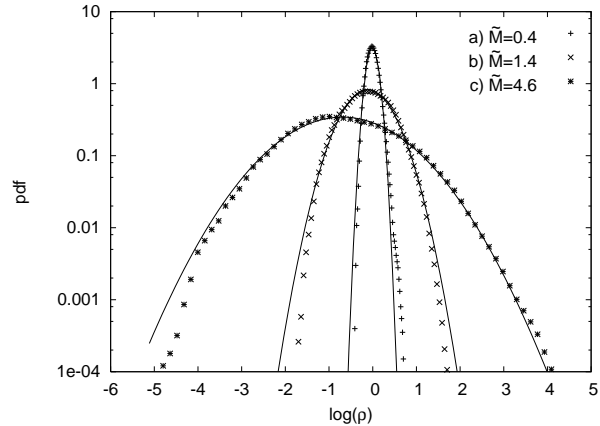}
  \caption{PDF of logarithm of mass density. Solid lines are normal distribution with
  mean and variance obtained from data.}
  \label{fig:pdf}
\end{figure}

\noindent
\textit{Eulerian statistics:} 
To characterize the flows we measured some specific values and show
velocity-spectra and mass-density PDFs in the turbulent regime (see
Figs. \ref{fig:spectra}, \ref{fig:pdf} and Table \ref{tab1}).
In fully developed isotropic turbulence, the energy brought into the
system at large scales should cascade through the scales and finally
dissipate at small scales, known as the Richardson cascade of energy.
In stationary turbulent flows, the energy injection equals the
dissipation. Therefore, the mean energy dissipation $\varepsilon$ can
be obtained from the mean energy input $\left<\partial_t E_k\right>$
of the driving force $\mathbf{k}$ averaged over several dynamical
times. Normalized to the mean kinetic energy in the box
$\varepsilon/E_{\mbox{{\tiny{kin}}}}$ we see in Table \ref{tab1} that
dissipation rises with the Mach number. In order to quantify the
Eulerian compressibility the dimensionless ratio $\mathcal{C}=\left <
  (\partial_i v_i)^2 \right > / \left < (\partial_i v_j)^2 \right >$
is measured which is zero for solenoidal and one for potential flows.
This quantity also shows a dependence of Mach number. Finally we give
an estimate for the integral scale Reynolds number $R=l_0 v_0/\nu$
defined by the integral length scale $l_0=2/3
E_{\mbox{{\tiny{kin}}}}^{3/2}/\varepsilon$.  Here, $v_0$ is the root
mean square velocity fluctuation, which in isothermal flows is the
same as the mean Mach number, and the dynamical viscosity $\nu$ is
given by $\nu=\eta^{4/3}\varepsilon^{1/3}$. The Kolmogorov dissipation
scale $\eta$ was estimated by the discretization scale $\eta=\Delta
x=L/\mathcal{N}$. Please note, that this is a conservative estimate
since in spectral simulations of incompressible flows the dissipation
scale $\eta$ is normally half the grid spacing.  Only for high Mach
number flows a Reynolds number can be reached which is comparable to
the one obtained from spectral simulations of incompressible flows
with the same resolution. The power spectrum of the velocity
fluctuations steepens with increasing Reynolds number reflecting the
occurrence of strong shocks. The slope of the spectrum in the inertial
range drops to a value, which is compatible to $P(k)\sim k^{-1.85}$
for run c. This is in agreement with the simulations of Kritsuk
\textit{et al.} \cite{kritsuk-norman-etal:2006} and Boldyrev
\textit{et al.}  \cite{boldyrev-nordlund-etal:2002}. Although the
latter is obtained from MHD simulations, it just reveals the dominance
of strong shocks that converge to Burgers-like spectra $\sim k^{-2}$
for even higher Mach-numbers.

\begin{table}[b]
\caption{\label{tab1}Characteristic quantities for the three runs}
\begin{ruledtabular}
\begin{tabular}{c c c c c}
Run  &   Mach &   $\mathcal{C}$ &   $\varepsilon/E_{\mbox{{\tiny{kin}}}}$ &   Reynolds number \\ \hline
a    &   0.4  &       0.07      &         0.31            &        250     \\
b    &   1.4  &       0.26      &         0.47            &        700     \\
c    &   4.6  &       0.50      &         0.95            &       1756         
\end{tabular}
\end{ruledtabular}
\end{table}

In the isothermal case the mass density PDF $R(\rho)$ should follow
a log-normal distribution as discussed in
\cite{passot-vasquez-semadeni:1998, nordlund-padoan:1999}.  In Fig.
\ref{fig:pdf} we compared the density PDF of the numerical simulations
with normal distributions with mean and variance obtained from the
numerical data which confirms the reasoning of
\cite{passot-vasquez-semadeni:1998, nordlund-padoan:1999}.

\noindent
\textit{Distribution of tracer particles:}
To obtain the PDF of the spatial distribution of the tracer particles
we divided the computational box into $n=40^3$ equally spaced
subdomains and counted the particles in them.  The number of boxes
with $k$ particles gave the PDF of the tracers as a discrete
distribution $T(k)$.  Since the particles were initially randomly
distributed, their PDF started from a Poissonian
\begin{displaymath}
  T_{init}(k)=P_{\lambda}(k)=\frac{\lambda^k}{k!}e^{-\lambda}
\end{displaymath}
with $\lambda=N/n$ and changed in time towards a stationary
distribution.

\begin{figure}[t]
  \centering
  \includegraphics[width=.95\linewidth]{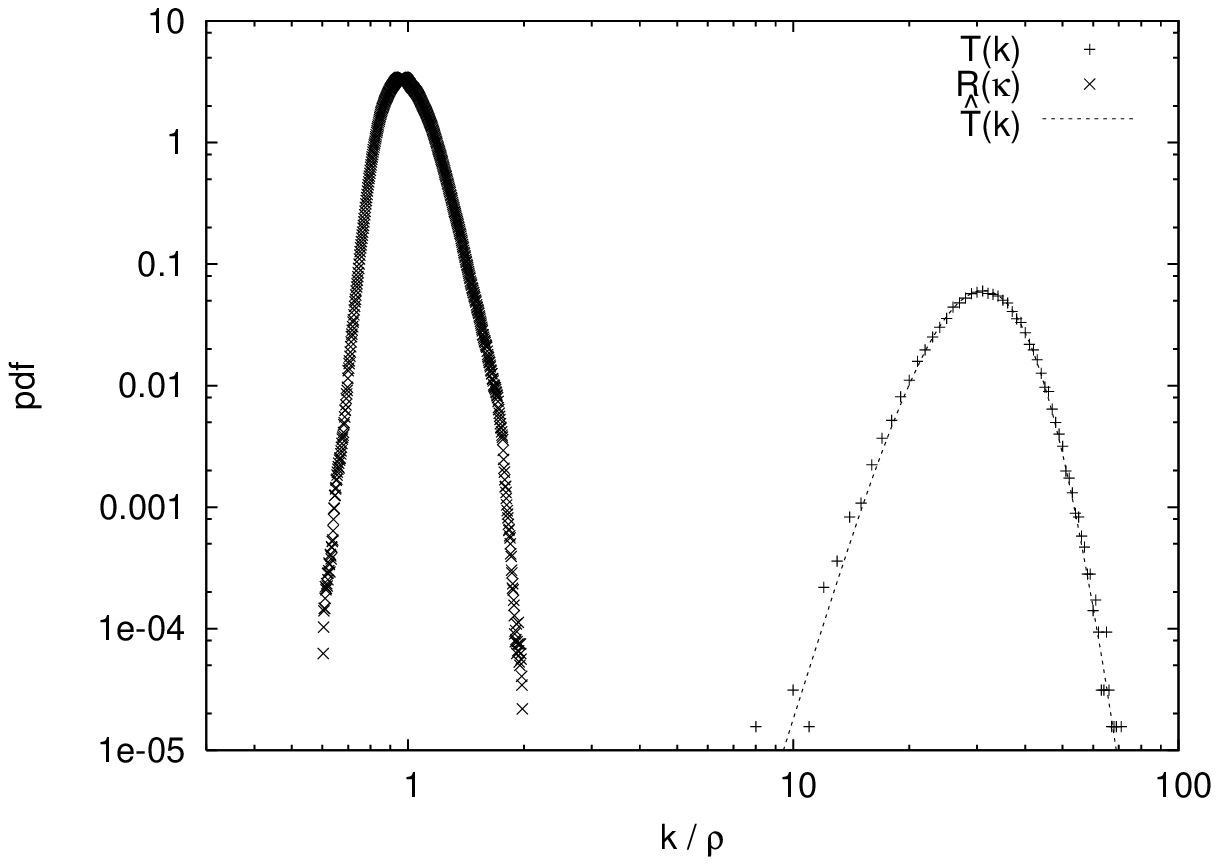}
  \includegraphics[width=.95\linewidth]{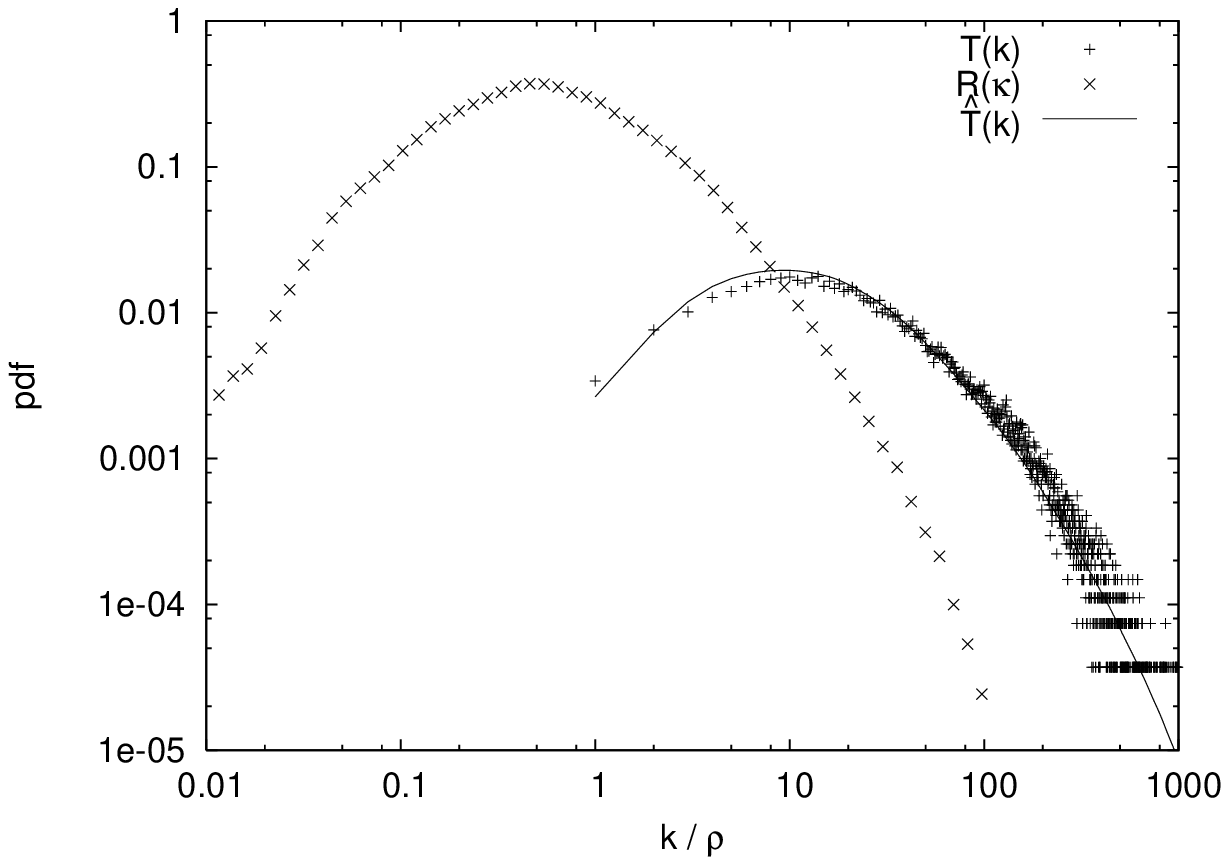}
  \caption{The convolution operator maps the mass density PDF $R(\rho)$ on the 
  particle distribution $\hat T(k)$ in good agreement with the measurement $T(k)$. 
  Top: run a, bottom: run c}
  \label{fig:particles}
\end{figure}

To understand the relation between the mass density PDF and the
particle distribution, we first consider an incompressible flow. The
PDF of $\rho \left(\vec x \right)=\rho_0 = \mbox{const}$ is
\begin{displaymath}
  R_{incomp}(\rho)=\delta(\rho-\rho_0)  \; .
\end{displaymath}
In this case the particle distribution stays the same for all times
and the expectation values $\mathbb{E}(R_{incomp})=\rho_0$ and
$\mathbb{E}(T)=\mathbb{E}(T_{init})=N/n$ are constant.  The
appropriate translation between the mass density PDF and the particle
distribution is a convolution which maps the $\delta$-peak to the
Poisson-distribution and matches the expected values:
\begin{eqnarray}
  \hat T (k) &=& (R \circledast P_{\lambda} ) (k) \nonumber \\
          :  &=& \int P_{\lambda=\frac{\mathbb{E}(T)}{\mathbb{E}(R)}\rho} \mathrm{d}\mu(\rho) \nonumber \\
             &=& \int P_{\lambda=\frac{\mathbb{E}(T)}{\mathbb{E}(R)}\rho}(k) R(\rho) \, \mathrm{d}\rho  \; .
\label{eq:convolution}
\end{eqnarray}
It is easy to see that this convolution works for the incompressible
case.  From that it is suggesting to take the same convolution for
compressible turbulence with the measured $R(\rho)$ instead of the
delta peak in the incompressible case. In this way the convolution
generates a discrete particle distribution from the continuous
distribution of density.  In Fig. \ref{fig:particles} we show the
PDFs for run a and c (The PDFs for run b agree similarly well, but
are omitted here). The solid lines $\hat T(k)$ are the result of the
convolution of the measured $R(\rho)$ with the Poisson distribution.
It reproduces in all cases nearly exactly the measured particle
distribution $T(k)$.

In order to obtain an analytical expression, we use the relation
between the deviation $\sigma$ of $\rho$ and the r.m.s. Mach number
$M$ for isothermal supersonic turbulence introduced by Passot and
Vazquez-Semadeni \cite{passot-vasquez-semadeni:1998} and Nordlund and Padoan
\cite{padoan-nordlund-etal:1997,nordlund-padoan:1999}:
\begin{displaymath}
  \sigma^2 = \beta^2 M^2  \: .
\end{displaymath}
This translates to the relation 
\begin{equation}
  \sigma_s^2  = \ln \left( {1 + \beta^2 \frac{{M^2 }}{{ \mathbb{E}^2 \left( R \right)}}} \right)
  \mbox{ .}
  \label{eq:sigma_sigma_s}
\end{equation}
of the standard deviation $\sigma_s$ of $\ln \rho$ and the Mach number $M$.

\begin{figure}[t]
  \centering
  \includegraphics[width=.95\linewidth]{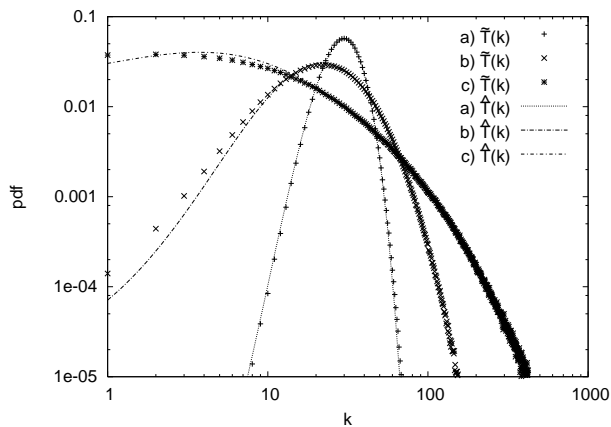}
  \caption{Particle distribution averaged over several time steps $\tilde T\left( k \right)$ and the theoretical distribution $\hat T \left( k \right)$ obtained from equation~(\ref{eq:lognorm_poisson}) with $\beta = 0.37$.}
  \label{fig:theory}
\end{figure}

If we convolve this lognormal distribution with a Poisson-distribution
using eqn. (\ref{eq:convolution}) we obtain
\begin{eqnarray}
  \label{eq:lognorm_poisson}
  \hat T(k)       &=& \frac{E^k}{\sqrt{2\pi}\sigma_s}\frac{1}{k!}\int\limits_{0}^{\infty}\rho^{k-1}\exp\left(-\frac{(\ln \rho - \mu)^2}{2\sigma_s^2}-E \rho \right ) \mathrm{d}\rho               \nonumber       \\
  \mbox{with}     \\
  E &=& \frac{\mathbb{E}(T)}{\mathbb{E}(R)}       \mbox{ .}               \nonumber
\end{eqnarray}
Hence, we can describe the particle-distribution with only one fit
parameter $\beta$.  In Fig. \ref{fig:theory} we show the agreement
between the particle-distribution and this theoretical PDF. This
figure was obtained with a value of $\beta \simeq 0.37$. This value lies
somewhere in between the value $\beta = 0.26$ of the recent simulations
of Kritsuk \textit{et al.} \cite{kritsuk-norman-etal:2007} and $\beta
= 0.5$ of the MHD simulations of Padoan \textit{et al.}
\cite{padoan-nordlund-etal:1997}.

\begin{figure}[t]
  \centering
  \includegraphics[width=.95\linewidth]{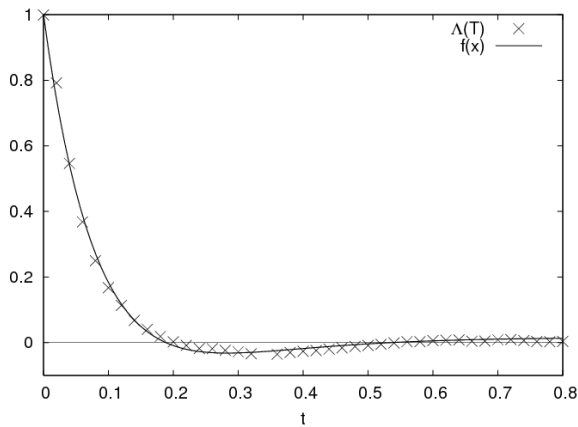}
  \caption{Transition of the particle distribution from the Poissonian
    to the $\hat T(k)$-distribution characterized by $\Lambda[T](t)$
    and compared to the exponential fit $f(t)$.
    \label{fig:structuring}}
\end{figure}

\noindent
\textit{Structuring time}
After this findings we can determine how long it takes for the
particles to form global structures.  This time $t_S$ is equivalent to
the time in which the particle PDF goes from the initial Poissonian to
the final distribution given in eqn. (\ref{eq:lognorm_poisson}). To
estimate this time we define an operator $\Lambda[T]$ on the particle
distribution $T$ which assigns the initial Poisson distribution
$T=P(k)$ to $1.0$ and the final stationary distribution $T=\hat T(k)$
to $0$:
\begin{displaymath}
  \sum\nolimits_k {\left( {(1-\Lambda[T]) \cdot \hat T\left( k \right) 
   + \Lambda[T] \cdot P\left( k \right) - T\left( k \right)} \right)^2 }  \to {\text{Min}}
\end{displaymath}
where the value $\Lambda[T]$ is obtained from the minimization
procedure.  A plot of the temporal development of $\Lambda[T]$ is
shown is Fig.~\ref{fig:structuring}.

To give an estimate for the time scale $t_S$ of the particle
clustering, an exponentially damped oscillating function is fitted to
the curve $\Lambda[T](t)$ with decay proportional to $\exp(-t/t_S)$.
In Table \ref{tab2} values for this structuring time $t_S$ as well as
the dynamical time $t_{dyn}$ are given. The comparison shows that the
structuring time is significantly shorter than the dynamical time,
although the ratio $t_S/t_{dyn}$ grows slightly with increasing Mach
number.

\begin{table}[bht]
\caption{\label{tab2}Comparison between the dynamical time $t_{dyn}$ and the structuring time $t_S$}
\begin{ruledtabular}
\begin{tabular}{c c c c c}
Run  &   Mach &   $t_{dyn}$   &   $t_S$                 \\ \hline
a    &   0.4  &      7.8      &    1.0      \\
b    &   1.4  &      2.2      &    0.33     \\
c    &   4.6  &     0.62      &    0.11         
\end{tabular}
\end{ruledtabular}
\end{table}

\noindent
\textit{Conclusions and outlook:}
In this Letter we have established a direct relationship between the
particle distribution and the PDF of the density fluctuations in
isothermal compressible turbulence. The analytical expression for the
particle distribution confirms the relation between the standard
deviation of $\ln \rho$ and the mean Mach number. The proportionality
constant $\beta$ was obtained from the particle distribution. The
Lagrangian viewpoint offers further studies on the formation of
density structures. In this paper, we studied only the low order
statistics.  The next natural step is to consider higher order moments
of the particle number in a given ball of radius $r$. A first step in
this direction for an unrealistic compressible Kraichnan flow was done
in Bec \textit{et al.} \cite{bec-gawedzki-etal:2004}.

\noindent
{\it Acknowledgments.---}
We'd like to acknowledge interesting discussions with R. Friedrich and J. Schumacher.
Computations were performed on a Linux-Opteron cluster supported by HBFG-108-291.
This work benefited from support through DFG-GK 1051 and DFG-SFB 591.
%

\end{document}